\documentclass[10pt,twocolumn]{IEEEtran}

 \usepackage{url}
\usepackage[cmex10]{amsmath}
\usepackage{amsfonts}
 \usepackage{theorem}
\usepackage[dvips]{epsfig}
\usepackage{epsfig}

 \usepackage{verbatim}
 \usepackage{amssymb}
\usepackage{amsbsy}
 \usepackage{setspace}
\usepackage{url}
\usepackage{subcaption}

\theoremstyle{plain}

\newtheorem{Proposition}{Proposition}

{\theorembodyfont{\rmfamily} }
{\theorembodyfont{\rmfamily} }
{\theorembodyfont{\rmfamily} }
{\theorembodyfont{\rmfamily} }
{\theorembodyfont{\rmfamily} }

\newcommand {\R}{\mathbb R}

\newcommand{\Int}{\operatorname{{\mathrm int}}}

\newcommand{\be}{\begin{equation}}
\newcommand{\ee}{\end{equation}}

\begin{document}

\title{ Sensitivity  of    mRNA Translation }

\author{Gilad Poker, Michael Margaliot and Tamir Tuller\thanks{The authors are with the School of Elec. Eng., Tel Aviv University, Israel 69978.
Corresponding author: Prof. Michael Margaliot, Email: michaelm@eng.tau.ac.il}}

\maketitle
\section{Abstract}
Using the dynamic mean-field approximation of the totally asymmetric simple exclusion process~(TASEP),
we investigate the effect of     small changes in
the initiation, exit, and elongation rates along the mRNA strand
on the steady state protein translation rate.
We focus on two special cases where exact closed-form
expressions  for the translation rate sensitivity can be derived. We discuss
the ramifications of our results
in the context of functional genomics, molecular evolution, and synthetic biology.

\section{Introduction}

During mRNA translation complex molecular machines called ribosomes attach to the~5' end
of the messenger RNA~(mRNA)   and then scan it
in a sequential manner.
At each elongation step, a
   nucleotide triplet (codon) is ``read''
    and the ribosome ``waits'' until  a
freely diffusing transfer RNA~(tRNA),  carrying  the
corresponding  amino-acid, binds to the ribosome.
The process ends when the ribosome reaches the~3' end of the mRNA,
detaches, and releases the chain of amino-acids that folds into a functioning protein~\cite{Alberts2002}.

Translation is a crucial step in gene expression, and it is becoming increasingly clear that
understanding this process is vital
 in order
to reveal how biological systems  develop, evolve, and function.
Indeed, mRNA translation is the most extensively regulated
step in mammals~\cite{Schwanhausser2011}, and a $100$-fold range
of translational efficiency
was detected between different   genes~\cite{Vasquez17012014,one_100_fold}.
This clearly has a strong effect on the protein abundance
that cannot be predicted by measuring mRNA abundances alone.

In this letter, we develop a new approach for studying the sensitivity of the   translation rate
with respect to changes in the genetic machinery. We show that our
analytical predictions agree well with recent experimental findings.
 
Two important dynamical aspects of the translation process
are: (1) certain codons are ``slower'' than others  dut to factors such as
low abundance of tRNA molecules with the corresponding
 anti-codon,   folding of the mRNA, and  interactions of the translated protein and the ribosome \cite{TullerGB2011};
and (2)~many ribosomes scan along the same mRNA chain in parallel and ``traffic jams''
can form behind a ``slow'' ribosome.

A mathematical model that encapsulates these properties is 
TASEP~\cite{TASEP_book,TASEP_tutorial_2011,MacDonald1968}.
In this model, particles move along a chain of~$n$ consecutive sites.
Each site can be either occupied by a particle or free.
A particle attaches to the first site with probability~$\alpha$ (but only if this site is free),
hops from site~$i$ to site~$i+1$ with probability~$\gamma_i$  (but only if  site~$i+1$ is free),
and hops from the last site of the chain  with probability~$\beta$.
In the homogeneous TASEP, all the transitions rates~$\gamma_i$ are assumed to be equal.
In the context of translation, the particles [chain] model the ribosomes [mRNA molecule].

The dynamic mean-field approximation of TASEP (see e.g., the excellent survey paper~\cite[p.~R345]{solvers_guide}),
sometimes called the \emph{ribosome flow model}~(RFM)~\cite{reuveni},
is a set of~$n$ ordinary differential equations:
\begin{align}\label{eq:rfm}
                    \dot{x}_1&=\lambda_0 (1-x_1) -\lambda_1 x_1(1-x_2), \nonumber \\
                    \dot{x}_2&=\lambda_{1} x_{1} (1-x_{2}) -\lambda_{2} x_{2} (1-x_3) , \nonumber \\
                    \dot{x}_3&=\lambda_{2} x_{ 2} (1-x_{3}) -\lambda_{3} x_{3} (1-x_4) , \nonumber \\
                             &\vdots \nonumber \\
                    \dot{x}_{n-1}&=\lambda_{n-2} x_{n-2} (1-x_{n-1}) -\lambda_{n-1} x_{n-1} (1-x_n), \nonumber \\
                    \dot{x}_n&=\lambda_{n-1}x_{n-1} (1-x_n) -\lambda_n x_n.
\end{align}
Here~$x_i(t) \in [0,1]$ is the normalized
occupancy level at site~$i$ at time~$t$,
$\lambda_i >0 $ is a parameter that controls
the transition rate from site~$i$   to the consecutive site~$i+1$.
 To explain this model, consider for example the
 equation
 \[
                  \dot{x}_1=\lambda_0 (1-x_1) -\lambda_1 x_1(1-x_2).
 \]
 The term~$\lambda_0 (1-x_1) $
 is the rate at which  ribosomes attach  to the beginning  of the chain.
 This is given by the product
 of~$\lambda_0$ (the initiation rate)
 with~$(1 - x_1)$. This means that as~$x_1$ increases, {\em i.e.}, as site~$1$ becomes fuller,
 the effective binding rate decreases.
 In particular, when~$x_1(t)=1$ the site is completely full
 and the effective binding rate is  zero.
 The term~$(1-x_1)$ thus
 reflects the simple exclusion principle of TASEP.
 The term~$\lambda_1 x_1(1-x_2)$ is the rate in which
  ribosomes move
 from site~$1$ to site~$2$. This is proportional to
 the occupancy level at site~$1$, and to~$(1-x_2)$
 representing again the simple exclusion principle.
 The symmetry between the~$x_i$ and~$(1-x_i)$ terms also preserves the
 particle-hole symmetry of TASEP.
 The term~$\lambda_n x_n$ describes the rate of
 ribosomes exiting from the last site, so
 $R(t):= \lambda_n x_n(t)$ is the protein translation
  rate at time~$t$.

Unlike TASEP, the RFM is a deterministic and continuous-time model.
Nevertheless, it has been shown that for the range of parameters that are relevant for
   translation RFM  and TASEP provide
highly  correlated predictions~\cite{reuveni}.

The state-space of the RFM is the unit cube~$C^n:=\{x\in \R^n:x_i \in[0,1]:i=1,\dots,n\}$.
For~$a \in C^n$, let~$x(t,a)$ denote the solution at time~$t$ of the RFM
emanating from~$x(0)=a$.
It is known~\cite{RFM_stability,RFM_entrain} that the dynamics admits a unique equilibrium~$e=e(\lambda_0,\dots,\lambda_n)$,
with~$e \in \Int(C^n)$, and that every trajectory of the RFM converges to~$e$, that is,
 $\lim_{t\to\infty} x(t,a)=e$ for all~$a\in C^n$.
In particular,~$R(t)$ converges to the \emph{steady-state
translation
 rate}~$R:=\lambda_n e_n$.

Substituting~$e$ for~$x$ in~\eqref{eq:rfm}  yields
\begin{align} \label{eq:rfm_ss}
       \lambda_0 (1- {e}_1) & = \lambda_1 {e}_1(1- {e}_2)\nonumber \\ 
                     & \vdots \nonumber \\
                    &= \lambda_{n-1} {e}_{n-1} (1- {e}_n) \nonumber \\
                    & =\lambda_n  {e}_n,
\end{align}
and this  gives
\begin{align}\label{eq:Rei}
	R=\lambda_i e_i(1-e_{i+1}), \quad i\in\{0,1,\dots,n\},
\end{align}
where~$e_0$ and~$e_{n+1}$ are  defined as~$1$ and~$0$, respectively.
From this it is possible to obtain an equation for~$R$
that includes a continued fraction. For example, for~$n=2$,
\eqref{eq:Rei} yields
\[
          1-\cfrac{R/ \lambda_0}{  1-\cfrac{R/\lambda_1} {1-R/ \lambda_2 }} = 0.
\]

It has been recently shown~\cite{rfm_max}
that the~$(n+2)\times(n+2)$ matrix:
\begin{align}\label{eq:rfm_matrix_A}
 \small
                A := \begin{bmatrix}
 0 &  \lambda_0^{-1/2}   & 0 &0 & \dots &0&0 \\
\lambda_0^{-1/2} & 0  & \lambda_1^{-1/2}   & 0  & \dots &0&0 \\
 0& \lambda_1^{-1/2} & 0 &  \lambda_2^{-1/2}    & \dots &0&0 \\
 & &&\vdots \\
 0& 0 & 0 & \dots &\lambda_{n-1}^{-1/2}  & 0& \lambda_{n }^{-1/2}     \\
 0& 0 & 0 & \dots &0 & \lambda_{n }^{-1/2}  & 0
 \end{bmatrix}
\end{align}
has real and distinct eigenvalues:
$
            \zeta_1< \zeta_2< \dots < \zeta_{n+2},
$
with
\be\label{eq:alp=r}
	 \zeta_{n+2} = R ^{-1/2} .
\ee
Furthermore, if we let~$q_i$, $i\in\{0,\dots,n\}$, denote the~$(i+1)\times(i+1)$ principal
minor of~$  (\zeta_{n+2} I- A)$ then
\be\label{eq:qi}
                q_i= R^{-(i+1)/2} e_i e_{i-1} \dots e_1 .
\ee

This   provides a powerful  linear-algebraic  framework for studying the steady-state
in the RFM and its dependence on the (generally inhomogeneous) entry, exit, and transition rates.
Note that since~$A$ is a (componentwise)
nonnegative matrix,~$\zeta_{n+2}$ is also the Perron root of~$A$, denoted~$\rho(A)$.

Let~$\R^{k}_{++}:= \{x\in \R^k:x_i>0,\; i=1,\dots,k\}  $. In~\cite{rfm_max},~\eqref{eq:alp=r} was applied  to prove that the mapping
$ (\lambda_0,\dots,\lambda_n) \to R$ is strictly concave on~$\R^{n+1}_{++}$.
This means that the problem of maximizing~$R$, subject to an affine constraint on the rates,
is a convex optimization problem. Maximizing~$R$, given the limited bimolecular budget,
 is important because it is known that translation
is one of the most energy consuming processes in the cell~\cite{Alberts2002}.
Also, maximizing  the translation rate is an important challenge in synthetic biology and specifically in heterologous gene expression  (see, for example, \cite{Gustafsson2004}).
For other  recent results on the analysis of the RFM using tools from systems and control theory, see, e.g.,~\cite{RFM_entrain,RFM_feedback,zarai_infi}.

The matrix~$A$ should not be confused with the
transition matrix used in the stochastic analysis of TASEP, as
that matrix decodes the transitions between all possible particle configurations and is thus
of dimensions~$2^n\times 2^n$. This limits its use to very short TASEPs only.

\begin{figure}[h]
	\centering
	\includegraphics[height=4.5cm,width=5.5cm]{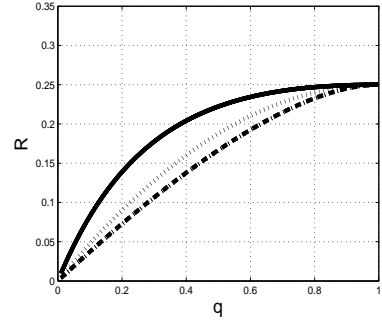}
	\caption{ Steady state translation rate~$R$ in a RFM with~$n=1000$ and rates~$\lambda_i=1$ except for a
configuration of slow rates with rate~$q$.
 Solid line: $\lambda_{500}=q$; Dotted line:~$\lambda_{500}=\lambda_{501}=q$;
Dashed line: $\lambda_{499}=\lambda_{500}=\lambda_{501}=q$.
As expected,~$R$ is a monotone function of~$q$. Note that a cluster of   consecutive slow sites
 considerably reduces~$R$.
 }
	\label{fig:config}
\end{figure}

Several interesting papers studied the effect of
  slow codon configurations  on the steady-state
  current in TASEP~\cite{PhysRevLett.93.198101,PhysRevE.74.031906,PhysRevE.76.051113,kolo_sense_1998}.
  The linear-algebraic representation of~$R$ provides a new, exact, and
 computationally  efficient approach  for studying this issue in the~RFM.
  For example, Fig.~\ref{fig:config}
  shows~$R$, computed via~\eqref{eq:alp=r}, for various slow rate
  configurations in an~RFM with~$n=1000$.

Here, however, we use~\eqref{eq:alp=r}
 to analyze  a different, yet related, notion, namely, the \emph{sensitivities}
 \[ s_i:=\frac{\partial}{\partial\lambda_i} R,\quad i=0,1,\dots,n .
 \]
 A relatively large value of~$s_k$
 indicates that the   rate~$\lambda_k$ has a strong  effect on the translation rate~$R$.
The advantage of our approach is that determining~$s_i$
   becomes an eigenvalue perturbation problem.
Since~$A$ is (componentwise) nonnegative and irreducible~\cite[Ch.~8]{matrx_ana}
there exists an  eigenvector~$v\in\R^{n+2}_{++}$ such that~$Av =\zeta_{n+2} v$.
By known results from linear algebra~\cite{magnus85},
\be\label{eq:der_pr}
                \frac{\partial }{\partial \lambda_i} \zeta_{n+2}= \frac{v' (\frac{d}{d\lambda_i}A) v }{v'v},
\ee
 so
\begin{align}\label{eq:derri}
                         s_i=  \frac{2R^{ 3/2} v_{i+1}v_{i+2} }{ \lambda_i^{3/2} v'v}.
\end{align}

 This provides a way to compute, in an  efficient  and numerically
 stable way, the sensitivities for large-scale RFMs using standard algorithms for computing
  the eigenvalues and eigenvectors of symmetric matrices.
\begin{figure*}[th]
	\centering
	\includegraphics[height=4.5cm,width=5.5cm]{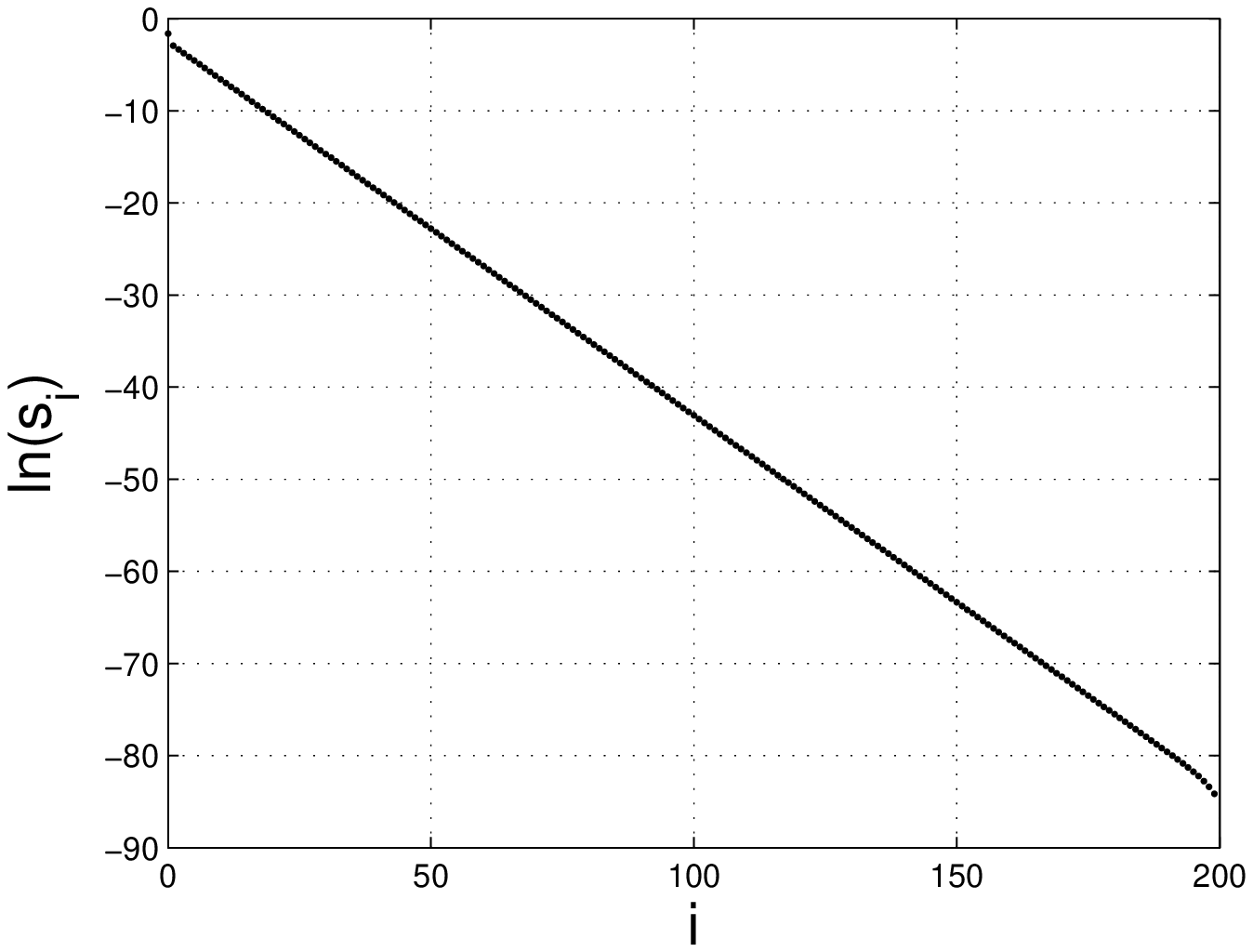}
	\includegraphics[height=4.5cm,width=5.5cm]{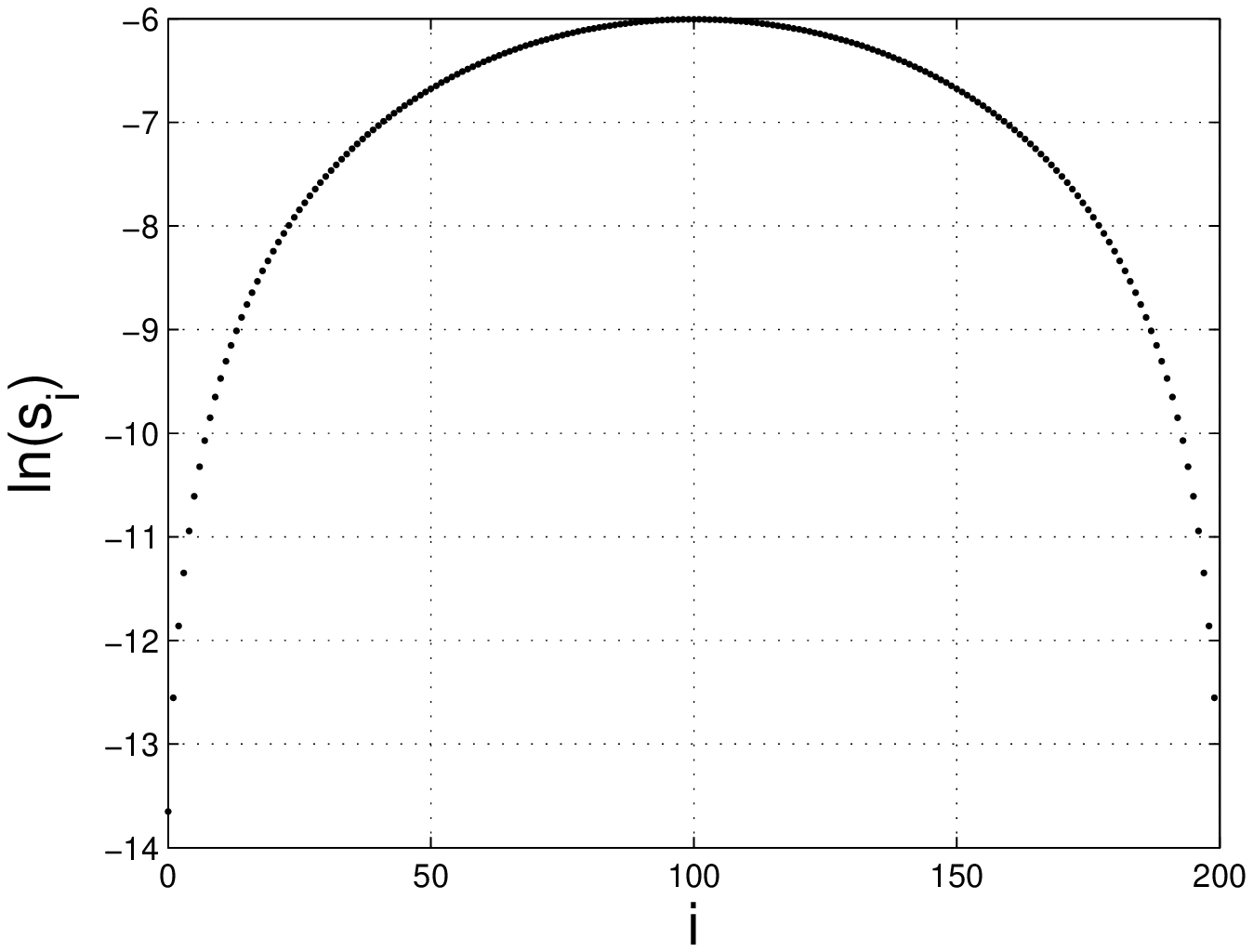}
	\includegraphics[height=4.5cm,width=5.5cm]{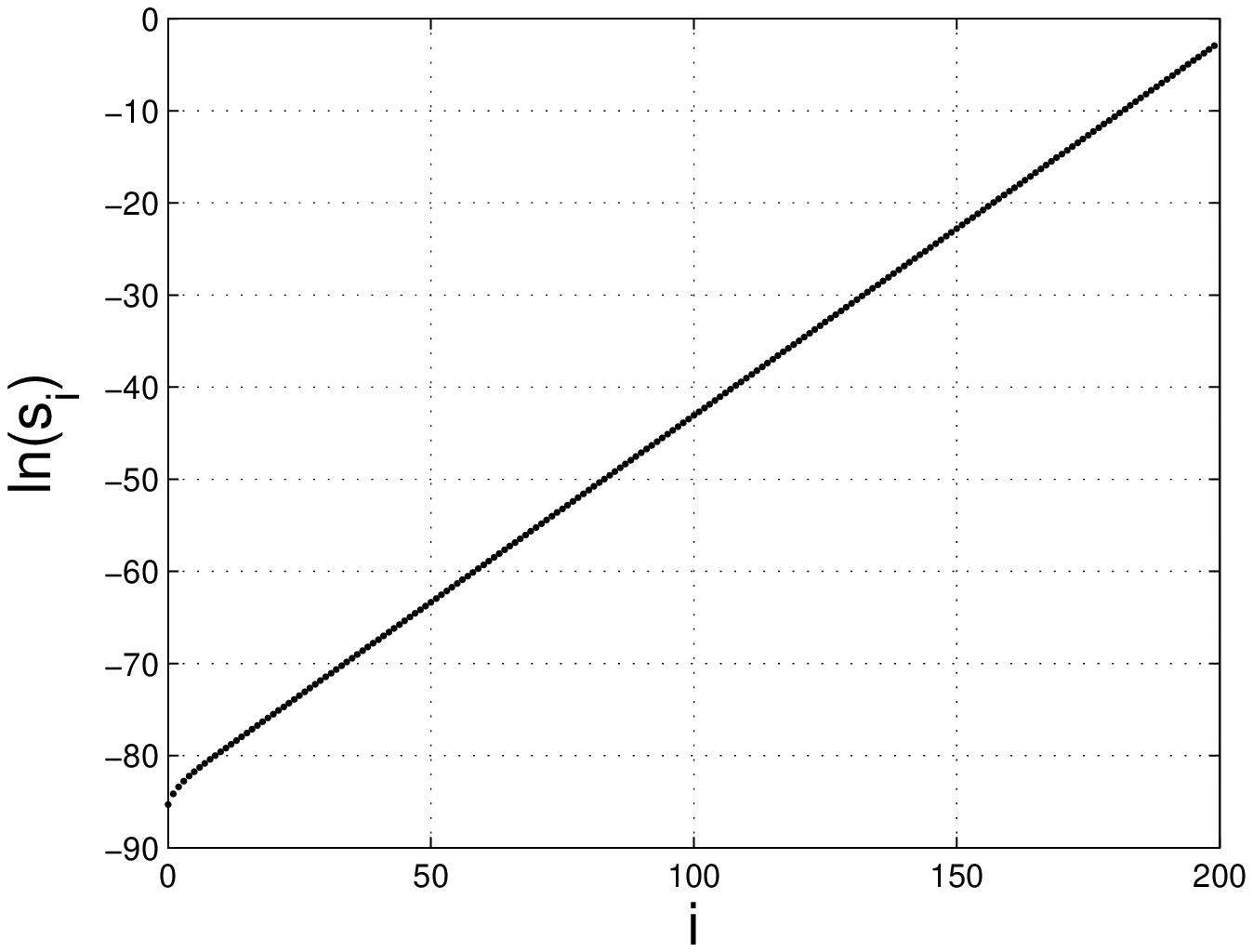}
	\caption{  $\ln(s_i)$ as a function of~$i$ in three HRFMs with~$n=200$.
Left: $\lambda_0=0.4$ and~$\lambda_{200}=1$.
Here~$\lambda_0$ is   rate limiting and thus the sensitivities of sites close to~$i=0$
are more important. Note that~$\ln(s_i)$ decays linearly with~$i$.
 Middle: $\lambda_0=\lambda_{200}=1$. Here the maximal sensitivity is with respect to~$\lambda_{n/2}$
  and it decays
 as we move towards the edges of the chain.
Right: $\lambda_0=1$ and~$\lambda_{200}=0.4$.}
	\label{fig:sens3}
\end{figure*}

 Fig.~\ref{fig:sens3}
 depicts~$\ln(s_i)$ as a function of~$i$ for three homogeneous RFMs~(HRFMs)
 ({\em i.e.}, $\lambda_i=1$ for all~$i \in\{1,\dots,n-1\}$),
 with size~$n=200$. The left sub-figure shows the case where~$\lambda_0=0.4$ and~$\lambda_{200}=1$,
 so~$\lambda_0$ is the rate limiting factor. The sensitivity~$s_0$
 is maximal and the sensitivities decrease as~$i$ increases.
 This regime describes the typical case in endogenous genes where initiation is the rate limiting factor~\cite{Malys2011}.
  Such genes indeed demonstrate
 selection for increased robustness to   transcription errors
in     ORF features that   affect the translation rate
  ({\em e.g.} mRNA folding and adaptation to the~tRNA pool)~\cite{TullerGB2011}.
  Similarly, our results may also explain   the evolutionary selection for unusual codon usage bias at the
   ORF 5' end~\cite{Tuller2014}.

 The right sub-figure shows the symmetric case where
 $\lambda_0=1$ and~$\lambda_{200}=0.4$.
 The middle sub-figure depicts the case where \emph{all} the~$\lambda_i$s are one.
  The plot then shows the TASEP  \emph{edge effect}~\cite{edge_tasep_2009}: the maximal
 sensitivity is in the center of the chain, and it decreases as
 we move toward the edges.
 This suggests that in order to maximize the translation rate
 in heterologous gene expression~\cite{Gustafsson2004}
 more attention should be devoted
to   tuning the codons in the middle of the coding sequence.


Pick~$i\in\{0,1,\dots,n\}$.
Since~$v_k>0$ for all~$k\in\{1,\dots,n+2\}$,
Eq.~\eqref{eq:derri} implies that~$s_i>0$, {\em i.e.}
an increase in any of the rates increases the steady state translation rate.
To determine an upper bound on~$s_i$,
perturb~$\lambda_i$ to~$\tilde \lambda_i : =\lambda_i +\epsilon $, with~$\epsilon>0$. This
  yields a perturbed matrix~$\tilde A$ that is identical to~$A$
except for
 entries $(i+1,i+2)$ and~$(i+2,i+1)$ that are:
\[
\tilde    \lambda_i  ^{-1/2} =
(\lambda_i + \epsilon )^{-1/2}
= \lambda_i^{-1/2} -  {\epsilon}  \lambda_i^{-3/2} /2 + o(\epsilon^2) .
\]
Thus,~$\tilde A=A+E $, where~$E$ is a matrix
with zero entries except for  entries $(i+1,i+2)$ and~$(i+2,i+1)$ that are
$ -\epsilon  \lambda_i^{-3/2}/2+ o(\epsilon^2)$.
By Weyl's inequality~\cite[Ch.~4]{matrx_ana},
\begin{align*}
\rho(\tilde{A}) & \geq
               \rho(A) - \epsilon  \lambda_i^{-3/2}/2+o(\epsilon^2).
\end{align*}
This yields~$ \frac{d\rho(A)}{d\lambda_i}\geq - \lambda_i^{-3/2}/2$,
so
$
s_i\leq \left(\frac{R}{\lambda_i}\right)^{3/2}$,
and since~$R\leq \lambda_i$ for all~$i$~\cite{reuveni},
$
s_i\leq 1 .
$
Thus, the maximal possible effect
of a small increase/decrease in any of the rates
is an increase/decrease of the same magnitude in the translation  rate. This  agrees
 with a recent experimental study on the
   change in protein abundance resulting
from perturbing  the codons of heterologous genes~\cite{Ben-Yehezkel2014}.
  In  this study, $25$ variants of the viral gene {\em HRSVgp04} were generated and the corresponding protein levels
   were measured in {\em S. cerevisiae}.
    In each variant {\em only}
     codons $41$-$80$ of the ORF were perturbed, without changing the encoded
     protein; thus, mRNA levels and translation initiation were
      expected to be identical in all variants.
  For each variant, the predicted change in the corresponding $\lambda_i$s ({\em i.e.} transition rates) was computed
  based on~\cite{Dana2014}. An average  change of $33.3 \%$ in the transition rate
  led to   a  $27.5 \%$ change in the protein levels.

\begin{figure}[h]
	\centering
	\includegraphics[height=4.5cm,width=5.5cm]{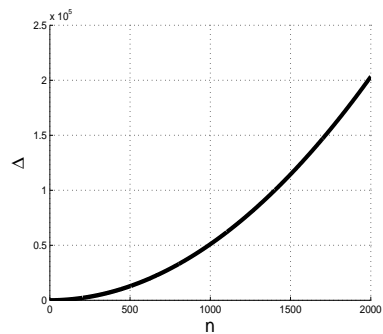}
	\caption{  $\Delta(n)$ as a function of~$n$.}
	\label{fig:delta}
\end{figure}

Below we focus on   two special cases  where
it is possible to obtain \emph{exact closed-form} expressions for the sensitivities.

\begin{figure*}[th]
	\centering
	\includegraphics[height=4.5cm,width=5.5cm]{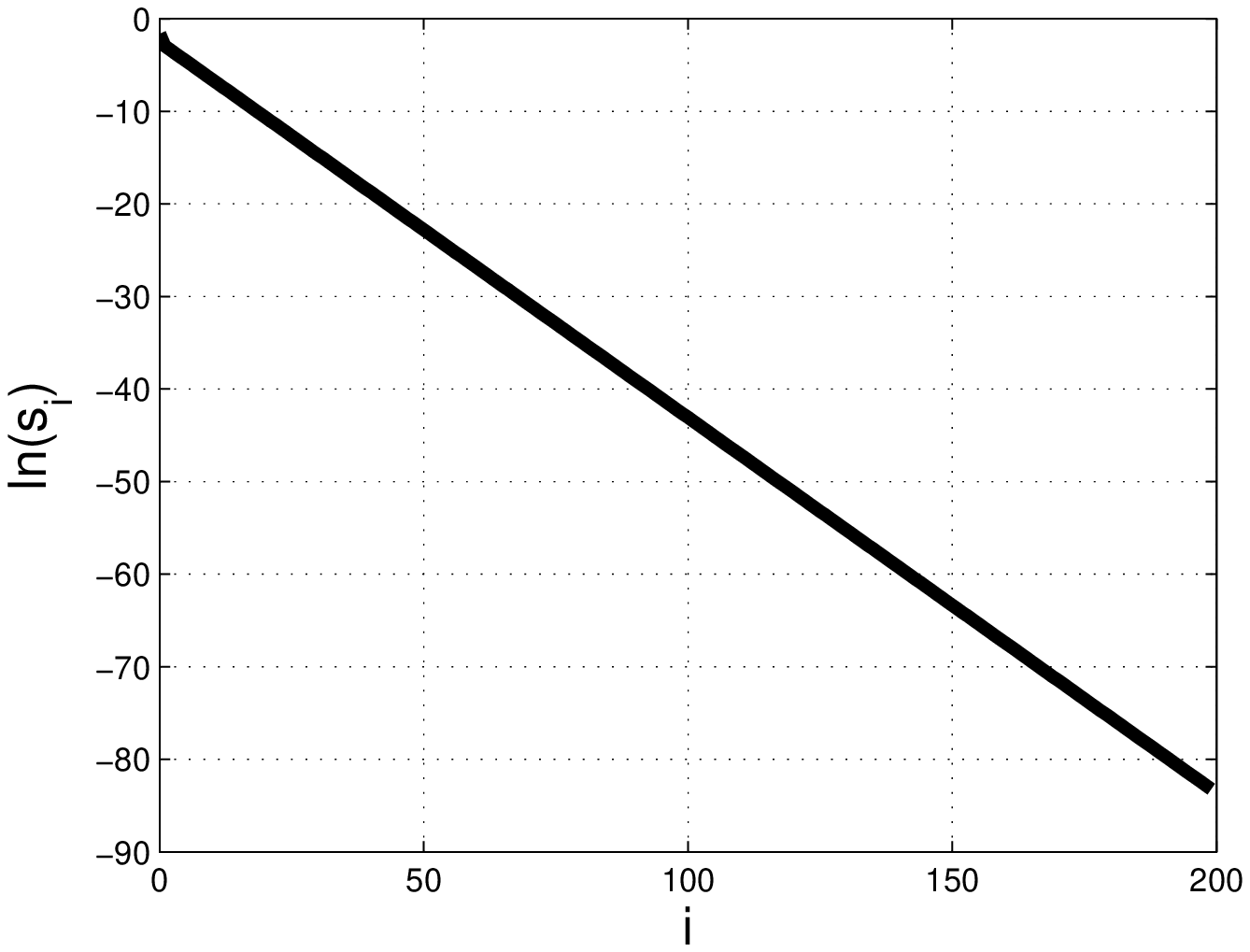}
	\includegraphics[height=4.5cm,width=5.5cm]{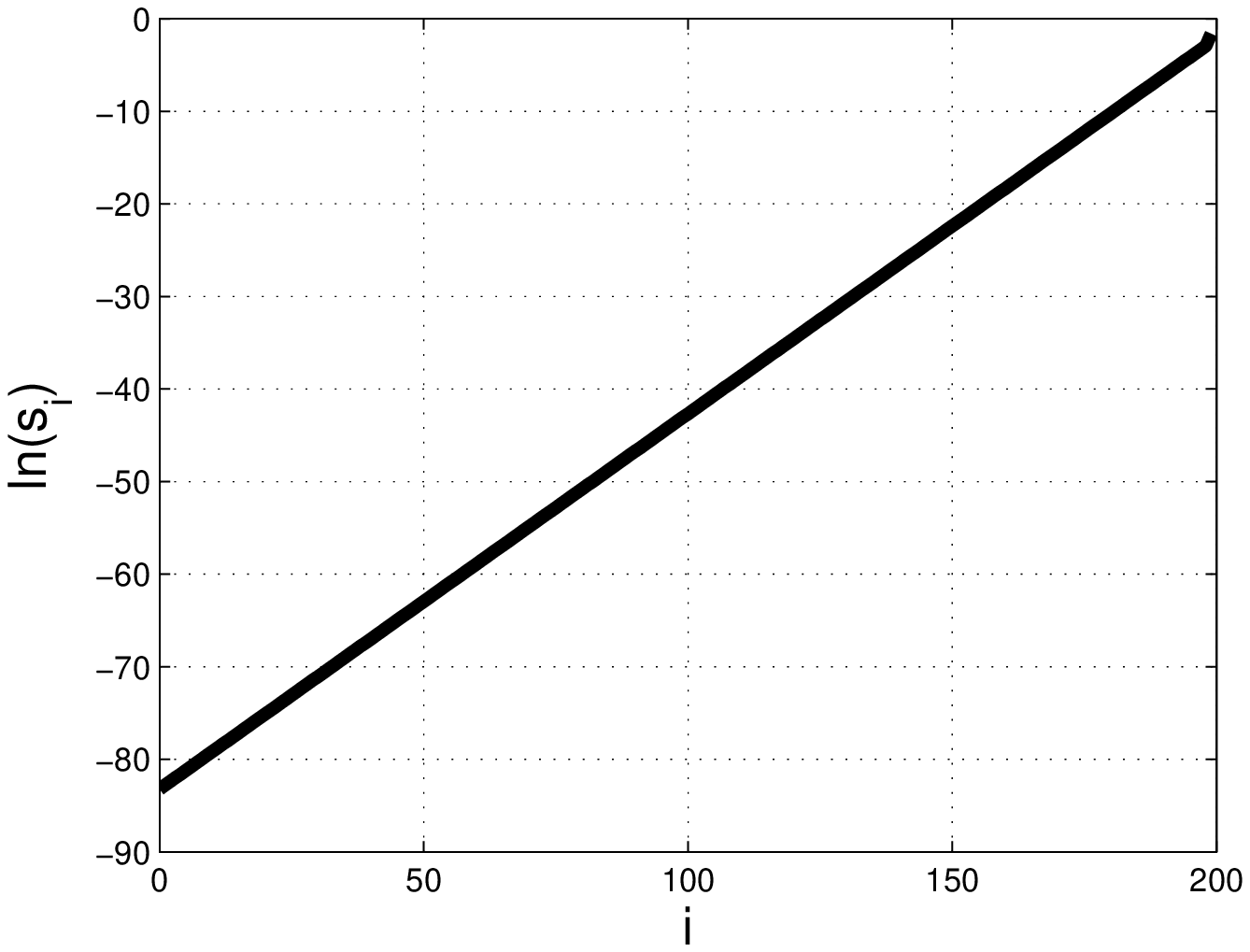}
	\caption{$\ln(s_i)$ as a function of~$i$ for two HSRFMs with~$n=200$.
Left: $\alpha=0.4$. Here~$1-\alpha=0.6$, so the entry rate is the limiting factor.
$s_{200}=-s_0<0$ is not shown.
Right: $\alpha=0.6$. Here~$1-\alpha=0.4$, so the exit rate is the limiting factor.
$s_0=-s_{200}<0$ is not shown.
}\label{fig:hsrfm}
\end{figure*}

\subsection{Totally homogeneous ribosome flow model~(THRFM)}
Suppose that~$\lambda_i=\lambda_c$ for \emph{all}~$i\in\{0,\dots,n\}$.
In other words, the initiation rate, exit rate,  and all   transition
rates are equal, with~$\lambda_c$ denoting their common value.
We refer to this case as the THRFM.
The matrix~$A$ in~\eqref{eq:rfm_matrix_A} then
becomes~$A=\lambda_c^{-1/2} B$, where
$B\in\R^{(n+2) \times(n+2)}$ is a tridiagonal Toeplitz matrix
with zeros on the main diagonal, and ones on the super- and sub-diagonal.
It is well-known~\cite{toeplitz_2013} that
the Perron root and Perron eigenvector of~$B$  are~$\rho(B)=2      \cos\left(\frac{\pi}{n+3} \right)$,
and ~$v(B)=\begin{bmatrix} \sin\left(\frac{ \pi}{n+3} \right) & \sin\left(\frac{ 2\pi}{n+3} \right) &\dots &
\sin\left(\frac{(n+2) \pi}{n+3} \right) \end{bmatrix}'$.
Therefore,~$\rho(A)=2\lambda_c^{-1/2}   \cos\left(\frac{\pi}{n+3} \right)$, and
\begin{align*}
	 R =\rho^{-2}(A)= \frac{\lambda_c}{4}     {\cos^{-2}\left(\frac{\pi}{n+3} \right)}.
\end{align*}
Substituting these values in~\eqref{eq:derri} and using the fact that
$\sum_{i=1}^{n+2}\sin^2(\frac{i\pi}{n+3}) = (n+3)/2$ shows that
the sensitivities   in the  THRFM are
                \be\label{eq:sen_thrfm}
                        s_i  =   \frac{\sin\left(\frac{i+1}{n+3}\pi\right)\sin\left(\frac{i+2}{n+3}\pi\right)}
                        {2(n+3)\cos^3\left(\frac{\pi}{n+3} \right)} ,\quad i=0,\dots,n.
                \ee
 This provides a closed-form expression for the graph shown
 in the middle plot of Fig.~\ref{fig:sens3}.

By~\eqref{eq:alp=r},    
$R(c\lambda_0,\dots,c\lambda_n)=c R( \lambda_0,\dots, \lambda_n) $ for all~$c>0$.
By  Euler's homogeneous function theorem,
$ R=\sum_{i=0}^n \lambda_i s_i $, and for the THRFM this gives
$  \sum_{i=0}^n s_i =\frac{1}{4}  {\cos^{-2}\left(\frac{\pi}{n+3} \right)} $.
Thus, $\lim_{n\to\infty} \sum_{i=0}^n s_i=1/4$.

 An exact      measure of the edge effect  is given by the ratio
 \begin{align*}
    \Delta (n)&:= \frac{s_{n/2}}{s_0} \\
     &=\left (  8\cos \left(  \frac{\pi}{n+3} \right) \sin^2 \left(  \frac{\pi/2}{n+3} \right)  \right)   ^{-1}
 \end{align*}
(see Fig.~\ref{fig:delta}). Note that
$\lim_{n\to \infty} \Delta(n)=\infty$. In other words, although all the sensitivities decay with~$n$ (see~\eqref{eq:sen_thrfm})
the edge effect actually becomes more prominent.
 One explanation for the edge effect here is that
 a site at the center of the chain has ``more neighbors'' than a site
located towards one of the edges of the chain. Thus, when all the rates are equal
the  protein translation  rate is more sensitive
to  the rates  at the center of the chain.

\subsection{Homogeneous and symmetric ribosome flow model~(HSRFM).}
Another  particular  case  where closed-form expressions for the~$s_i$s
can be derived
is when
$
            \lambda_1=\lambda_2=\dots=\lambda_{n-1}:=\lambda_c,
$
and
\be\label{eq:constraint}
\lambda_n=\lambda_c-\lambda_0.
\ee
We refer to this case as the~HSRFM.
The matrix~$A$ in~\eqref{eq:rfm_matrix_A} then becomes~$A= \lambda_c^{-1/2}C$, where
\begin{align*}\label{eq:accA}
\small
C:=                  \begin{bmatrix}
 0 &  \alpha^{-1/2}   & 0   & \dots &0&0 \\
\alpha^{-1/2} & 0  & 1     & \dots &0&0 \\
 0& 1 & 0    & \dots &0&0 \\
 & & \vdots \\
 0& 0 & \dots   &1  & 0& (1-\alpha) ^{-1/2}     \\
 0& 0 & \dots   &0 & (1-\alpha) ^{-1/2}  & 0
 \end{bmatrix},
\end{align*}
with $\alpha:=\lambda_0/\lambda_c$. For~$\alpha<1$,
$\rho(C) =(\alpha(1-\alpha))^{-1/2} $,
and~$v(C)$ is given by
  \be\label{eq:vecvv}
v_i  = \begin{cases}
           1 ,        & i=1,\\
            \mu^{(i-1)/2} \alpha^{-1/2}    ,        & 2\leq i  \leq n+1,\\
             \mu^{ n/2 }    ,        &  i=n+2,
\end{cases}
\ee
 where~$\mu:=\alpha/(1-\alpha)=\lambda_0/\lambda_n$.
Note that if~$\alpha=1/2$ then
$v'v = 2(1+n)$, and otherwise~$v'v=   \frac{2(1-\mu^{n+1})}{1-\mu}    $.
Thus,~$R=\rho^{-2}(A)= \lambda_c \alpha(1-\alpha)$.
The sensitivities   in the  HSRFM can now be determined.\footnote{Here~$s_i$, $1\leq i \leq n-1$, were calculated using~\eqref{eq:derri}.
Due to  the additional coupling  in~\eqref{eq:constraint}, $s_0$ and~$s_n$ cannot be computed using~\eqref{eq:derri}, so
we used~\eqref{eq:der_pr}.}
 If~$\alpha= 1/2$ then
\[
                        s_i  =  \begin{cases}
             0   ,        & i=0 \text{ or } i=n,\\
              \frac{1}{ 4(n+1) }  ,        & 1\leq i  \leq n-1,
\end{cases}
\]
and otherwise,
\[
                        s_i  =  \begin{cases}
            1-2\alpha  ,        & i=0,\\
                \alpha (1-2\alpha)  \frac{\mu^i }{1-\mu^{n+1}}   ,        & 1\leq i  \leq n-1,\\
             2\alpha-1,        &  i=n.
\end{cases}
\]
This can be interpreted as follows.
If~$\lambda_0<\lambda_n$ (so~$\alpha<1/2$ and~$\mu<1$) then
$
    \frac{         s_{i+1} }{       s_{i } }=\mu
$
for all~$i\in\{ 1,\dots,n-2 \}$, i.e. inside the chain the sensitivity strictly
decreases  with~$i$.
 This is reasonable, as in this case   the initiation
 rate is the rate limiting factor.
 Note that~$s_n<0$. This is due to~\eqref{eq:constraint}, as increasing~$\lambda_n$ means decreasing
 $\lambda_0$, and since this is the rate limiting factor, this  decreases~$R$.
 The case~$\lambda_0 > \lambda_n$ is symmetric.

Fig.~\ref{fig:hsrfm} depicts the~$s_i$s for~$n=200$ and two values of~$\alpha$.
Comparing this to Fig.~\ref{fig:sens3}
shows that the explicit equations for the HSRFM actually provide very good approximations
to the general behavior of the HRFM in the case where either the entry rate or the exit rate are the rate limiting factors.

Summarizing,
  steady-state properties
of the dynamic mean-field approximation of TASEP
can be represented in a linear-algebraic form.  Using this representation,
we studied the sensitivity of the steady-state translation rate to
 perturbations in  the initiation, transition, and exit rates.
In this context, the problem reduces to the sensitivity of the Perron root
of a symmetric, nonnegative, tridiagonal matrix. This leads to:
(1)~efficient numerical computation of the sensitivities that is thus applicable
for large-scale models; and (2)~exact, closed-form expressions for the sensitivities in some special,
yet important,  cases.

\bibliographystyle{IEEEtranS}

\bibliography{RFM_bibl_for_rfm_sense}

\end{document}